\documentclass[11pt]{article}
\usepackage[T1]{fontenc}
\usepackage{lmodern,amsmath,amssymb,amsthm,mathtools}
\usepackage[margin=1in]{geometry}
\usepackage{microtype}
\usepackage{array,booktabs,tabularx}
\usepackage[hidelinks]{hyperref}
\allowdisplaybreaks[1]
\newtheorem{theorem}{Theorem}[section]
\newtheorem{lemma}[theorem]{Lemma}

\theoremstyle{definition}

\theoremstyle{remark}
\newtheorem{remark}[theorem]{Remark}
\newcommand{\F}{\mathbb F}
\newcommand{\Z}{\mathbb Z}
\newcommand{\PG}{\operatorname{PG}}
\newcommand{\wt}{\operatorname{wt}}
\newcommand{\supp}{\operatorname{supp}}
\newcommand{\Span}{\operatorname{span}}

\newcommand{\E}{\mathbb E}
\newcommand{\Prob}{\mathbb P}
\DeclareMathOperator{\ord}{ord}
\numberwithin{equation}{section}
\title{A Hyperplane Bound and Ward's Divisibility Conjecture}
\author{Guanghui~Zhang \footnote{E-mail address: {\it zghui@squ.edu.cn.}}}

\date{\small (School of Mathematics and Physics, Suqian University, Suqian 223800, China)}
\hypersetup{pdftitle={A hyperplane bound and Ward's divisibility conjecture}}

\begin{document}
\maketitle

\begin{abstract}
Let $q=p^f$, where $p$ is prime. We prove that every $q$-ary
linear code meeting the Griesmer bound, with positive dimension and
minimum distance $d\ge1$, is
$p^{\max\{0,\nu_p(d)-f+1\}}$-divisible, as asserted in Ward's
conjecture. Arbitrary projective column multiplicities are allowed.
The geometric ingredient is a bound independent of dimension: a
nonzero scalar-invariant function $F:\F_q^k\to\F_p$, with $F(0)=0$
and prime-field algebraic degree at most $(f+s)(p-1)$, vanishes
identically on at most $p^s(q-1)$ linear hyperplanes when $0\le s<f$.
A truncated digit-sum inequality forces selected coefficient
polynomials to be homogeneous of ordinary degree less than $q$.
Weighted moments on a projective line then bound their linear
factors, while a second-moment argument handles the complementary
support configuration. An elementary finite-difference calculation
supplies the degree bound for divided weight residues. Endpoint
shortenings prove the basic divisibility layers; endpoint lifting,
a surjective linear functional, and residual-code induction extend
the argument to every valuation. We include the proofs of the
classical ingredients, identify their sources, and distinguish the
present argument from an earlier reported unpublished proof.

\medskip
\textbf{MSC:} 94B05; 94B27; 51E20.
	
\textbf{Keywords:} Griesmer codes; Ward's divisibility conjecture; prime-field algebraic degree; hyperplane bounds; finite differences.

\end{abstract}

\section{Introduction}

\subsection{Background and motivation}

Throughout the paper,
\[
 q=p^f,\qquad p\ \text{prime},\qquad f\ge1.
\]
An $[n,k,d]_q$ linear code is a $k$-dimensional subspace
$C\le\F_q^n$ with
\[
 \supp(c)=\{i\in\{1,\ldots,n\}:c_i\ne0\},\qquad
 \wt(c)=|\supp(c)|,\qquad
 d=\min_{c\in C\setminus\{0\}}\wt(c).
\]
We assume $k\ge1$ and $d\ge1$. For positive integers $k,d$, define
the Griesmer sum by
\begin{equation}\label{eq:griesmer-sum}
 g_q(k,d)=\sum_{i=0}^{k-1}\left\lceil\frac{d}{q^i}\right\rceil.
\end{equation}
The Griesmer bound is
\[
 n\ge g_q(k,d).
\]
A code satisfying $n=g_q(k,d)$ is called a \emph{Griesmer code}.
Griesmer proved the binary bound \cite{Griesmer}, and Solomon and
Stiffler extended it to every finite field \cite{SolomonStiffler}.
We recall the residual-code proof in Lemma~\ref{lem:residual}.

Equality in this length bound imposes restrictions on the entire
weight set. A positive integer $\Delta$ is a \emph{divisor} of $C$,
or $C$ is \emph{$\Delta$-divisible}, if
\[
 \wt(C):=\{\wt(c):c\in C\}\subseteq\Delta\Z.
\]
For a positive integer $m$, write
\[
 \nu_p(m)=\max\{a\in\Z_{\ge0}:p^a\mid m\}.
\]
Ward's conjecture \cite[Conjecture 10]{Ward2001} states that
\begin{equation}\label{eq:ward-original}
 p^e\mid d,\quad e\ge f
 \quad\Longrightarrow\quad
 \wt(C)\subseteq p^{e-f+1}\Z.
\end{equation}
The conjecture thus relates the prime-power divisibility of the
minimum distance to that of every codeword weight. We prove
\eqref{eq:ward-original} in all dimensions, allowing arbitrary
column multiplicities.

\subsection{Related work}

Earlier divisibility theorems provide several cases of Ward's
assertion. We recall these results and the geometric and
polynomial methods that enter the proof.

\begin{enumerate}
\item
\emph{Binary and prime-field Griesmer codes.}
Dodunekov and Manev proved the binary divisibility theorem
\cite[Lemma 3]{DodunekovManev}; Ward extended it to every prime field
\cite{Ward1998}:
\[
 C\ \text{Griesmer over }\F_p,\quad p^e\mid d
 \quad\Longrightarrow\quad p^e\mid\wt(c)
 \quad(c\in C).
\]
Ward also proved the first extension-field layer and the full
quaternary assertion
\cite[Proposition 13 and Theorem 14]{Ward2001}:
\[
 q\mid d\ \Longrightarrow\ p\mid\wt(c),\qquad
 q=4,\ 2^e\mid d,\ e\ge2
 \ \Longrightarrow\ 2^{e-1}\mid\wt(c).
\]
Consequently, the prime-field case, the first layer, and the
quaternary case are earlier results.

\item
\emph{General bounds of Deng, Huang, and Xiang.}
Their two divisibility theorems
\cite[Theorems 1.13 and 1.14]{DHX} yield, for $e=\nu_p(d)$,
\begin{equation}\label{eq:dhx-comparison}
 \wt(C)\subseteq p^{E_1}\Z,\qquad
 \wt(C)\subseteq p^{E_2}\Z,\qquad
 E_1=\left\lfloor\frac e f\right\rfloor,\quad
 E_2=\max\{0,e-(f-1)(q-2)\}.
\end{equation}
The theorem numbers refer to the journal article; the corresponding
results are Theorems 1.7 and 1.8 of arXiv~v1. Their work also
develops bases adapted to Griesmer subcodes, which enter the
structural theory recalled below. The exponent in our conclusion is
\[
 E_{\mathrm W}=\max\{0,e-f+1\}.
\]
The formulas are compared in Table~\ref{tab:divisibility-comparison}.
For example, $(q,e)=(8,5)$ gives $(E_1,E_2,E_{\mathrm W})=(1,0,3)$,
and $(q,e)=(9,3)$ gives $(1,0,2)$.
These are comparisons of divisibility guarantees for the indicated
parameter regimes.

\item
\emph{Geometric criteria with additional hypotheses.}
Landjev and Rousseva
\cite[Theorem 2]{LR} consider a multiset of points in $\PG(r,q)$,
$r\ge2$, with a fixed maximum point multiplicity. Assuming that
all lines through maximum-multiplicity points have one common
multiplicity, more than $(q-1)p^{t-1}$ such points force
$p^t$-divisibility, for $1\le t\le(r-1)f$.
Here divisibility of a multiset of total size $n$ means that every
hyperplane multiplicity is congruent to $n$ modulo $p^t$.
The common-line-multiplicity assumption is part of their theorem;
it is not assumed in Theorem~\ref{thm:hyperplane}.

Their later conference paper \cite[Theorem 5]{LR2026} gives
\[
 q^h\mid d,\quad h\ge2,\quad
 \left\lceil\frac{d}{q^{k-1}}\right\rceil=1
 \quad\Longrightarrow\quad
 p^2q^{h-2}\mid\wt(c)
 \quad(c\in C)
\]
for Griesmer codes. The displayed ceiling condition places the
result in the projective case, by the maximum-multiplicity bound.
For the same assumption $q^h\mid d$, our conclusion gives the
divisor $pq^{h-1}$ and also allows repeated columns.

\item
\emph{Polynomial degree and hyperplane counting.}
Two functional-degree results supply the polynomial framework.
Aichinger and Moosbauer's finite-field degree identity
\cite[Theorem 10.3, journal version]{AichingerMoosbauer} identifies
prime-field degree with the digit sum of field exponents.
Clark and Schauz's general theorem
\cite[Theorem 4.9(c)]{ClarkSchauz} gives the finite-difference
bound that we specialize below to functions
\[
 (\Z/p\Z)^f\longrightarrow\Z/p^{s+1}\Z.
\]
We give direct proofs of both specializations.
Haramaty, Ron-Zewi, and Sudan \cite[Lemma 4.8]{HRS} bound zero
hyperplanes for ordinary-degree polynomials using support
estimates and a second moment. Lemma~\ref{lem:incidence} uses
the same counting mechanism in projective form. The coefficient
argument in Section~\ref{sec:coefficients} addresses the additional
issue here: the available degree bound is a prime-field degree
bound, whereas the linear-factor count needs ordinary degree
strictly less than $q$.
\end{enumerate}

\begin{table}[tbp]
\centering
\caption{Guaranteed exponents $E$ in the divisor $p^E$ for a
Griesmer code over $\F_{p^f}$, where $e=\nu_p(d)$.}
\label{tab:divisibility-comparison}
\renewcommand{\arraystretch}{1.25}
\begin{tabularx}{\textwidth}{@{}Xl@{}}
\toprule
Result & Exponent $E$\\
\midrule
Deng--Huang--Xiang \cite[Theorem 1.13]{DHX}
 & $\lfloor e/f\rfloor$\\
Deng--Huang--Xiang \cite[Theorem 1.14]{DHX}
 & $\max\{0,e-(f-1)(q-2)\}$\\
Ward's asserted divisor, proved in Theorem~\ref{thm:ward}
 & $\max\{0,e-f+1\}$\\
\bottomrule
\end{tabularx}
\end{table}

The passage from the geometric bound to code divisibility uses
classical structural properties of Griesmer codes. Ward's endpoint
count and lifting lemmas \cite[Lemmas 11 and 12]{Ward2001} already
appear in his quaternary proof alongside residual induction and a
projective-line weight identity \cite[Theorem 14]{Ward2001}.
Residual and shortening equalities are developed further in
\cite{DHX}, with generalized-weight counterparts discussed in
\cite{DengGeneralized}. We include the elementary proofs needed
here. The final argument combines this framework with the
hyperplane theorem and a surjective linear functional over
an arbitrary finite field.

The historical record also includes a reported unpublished proof.
Bouyukliev \cite[Theorem 26]{Bouy2009} states the full assertion
\eqref{eq:ward-original} and attributes its general proof to
Xiaoyu Liu's 2006 preprint \emph{Divisibility of Griesmer Codes}.
We have not located that preprint and therefore cannot compare
its argument with ours. Liu's publicly available 2006 dissertation,
\emph{On Divisible Codes over Finite Fields}
\cite[p.~7]{LiuThesis}, is a different document and describes
the general assertion as a conjecture. The later papers
\cite{DHX} likewise present the extension-field assertion
as a conjecture or an open problem. These differing reports do not
by themselves establish priority. No result from the unavailable
preprint is used below.

\subsection{Our contribution and main results}

We state the divisibility theorem in a form that includes the
valuations for which the divisor is $1$.

\begin{theorem}[Ward's divisibility assertion]\label{thm:ward}
Let $C$ be a $q$-ary Griesmer code of positive dimension and
minimum distance $d\ge1$, where $q=p^f$. Then
\[
 \wt(C)\subseteq p^{\max\{0,\nu_p(d)-f+1\}}\Z.
\]
The statement allows repeated projective columns and imposes no
additional restriction on dimension or minimum distance.
\end{theorem}

The proof rests on a geometric bound for functions of small
prime-field degree. To state it, we specify the degree conventions
and projective notation. If $V=\F_q^k$, then
$\PG(V)=\PG(k-1,q)$ is the set of one-dimensional $\F_q$-subspaces
of $V$. A point is written
\[
 \langle v\rangle=\{\lambda v:\lambda\in\F_q\},
 \qquad v\in V\setminus\{0\}.
\]
A projective line is the projectivization of a two-dimensional
linear subspace. A linear hyperplane is a codimension-one
$\F_q$-subspace
\[
 H_\lambda=\ker\lambda,\qquad
 \lambda\in V^*\setminus\{0\},\qquad
 V^*=\operatorname{Hom}_{\F_q}(V,\F_q).
\]
The equations $\lambda$ and $\lambda'$ define the same hyperplane
exactly when $\lambda'=a\lambda$ for some $a\in\F_q^*$.

Choose a basis $\varepsilon_1,\ldots,\varepsilon_f$ of $\F_q$
over $\F_p$ and write
\[
 x_h=\sum_{j=1}^f z_{h,j}\varepsilon_j,\qquad
 z_{h,j}\in\F_p,\qquad 1\le h\le k.
\]
Every function $F:V\to\F_p$ has a unique reduced polynomial in
these $fk$ variables, with each exponent at most $p-1$.
Its total degree is denoted by $\deg_pF$ and called its
\emph{prime-field algebraic degree}.
We use the convention $\deg_p0=-\infty$.
Ordinary degree, denoted by $\deg$, refers instead to total degree
in $\F_q$-coordinate variables. Section~\ref{sec:coefficients}
proves the relation between the two and the independence of
$\deg_pF$ from the chosen prime-field basis.

A function $F:V\to\F_p$ is \emph{scalar invariant} if
\[
 F(av)=F(v)\qquad(a\in\F_q^*,\ v\in V).
\]
For such a function, $F(\langle v\rangle):=F(v)$ is well defined.
When $F(0)=0$, put
\begin{equation}\label{eq:zero-hyperplanes}
 \mathcal Z(F)=\{H\le V:\dim_{\F_q}H=k-1,\ F|_H\equiv0\}.
\end{equation}
The members of $\mathcal Z(F)$ are its
\emph{whole-zero hyperplanes}; they are counted as subspaces,
not as choices of defining forms.

\begin{theorem}[Dimension-free hyperplane bound]\label{thm:hyperplane}
Let $k\ge2$, let $0\le s<f$ be an integer, and let
$F:\F_q^k\to\F_p$ be a nonzero scalar-invariant function satisfying
\[
 F(0)=0,\qquad \deg_pF\le(f+s)(p-1).
\]
Then
\[
 |\mathcal Z(F)|\le p^s(q-1).
\]
\end{theorem}

The proof of Theorem~\ref{thm:hyperplane} begins with the
intersection of the support with projective lines. A short
nonempty intersection yields nonzero weighted moments. The
digit-cost inequality converts each suitable moment into a
nonzero homogeneous coefficient polynomial whose linear factors
count whole-zero hyperplanes. If no line has such an intersection,
the support is large enough for an incidence estimate to give
the bound. At $s=0$, the constant $q-1$ is attained by
$x_1^{q-1}-x_2^{q-1}$; see Remark~\ref{rem:sharp-first}.

For a code with generator matrix $G$, we apply this bound to
the divided weight residues
\[
 F_s(u)=\frac{\wt(uG)}{p^s}\pmod p,
\]
defined once $p^s$ divides all codeword weights.
Lemma~\ref{lem:degree} gives
$\deg_pF_s\le(f+s)(p-1)$.
For a minimum-dimensional counterexample to the next
divisibility layer, every endpoint shortening has that layer.
Its defining hyperplane therefore lies in $\mathcal Z(F_s)$.
The resulting contradiction is
\begin{equation}\label{eq:intro-contradiction}
 p^sq=p^{f+s}\le |E|
 \le|\mathcal Z(F_s)|\le p^s(q-1),
\end{equation}
where $E$ is the endpoint set and the lower bound follows from
Ward's endpoint count.

This proves, in succession, the implications
\begin{equation}\label{eq:intro-window}
 p^{f+s}\mid d
 \quad\Longrightarrow\quad
 \wt(C)\subseteq p^{s+1}\Z,
 \qquad s=0,\ldots,f-1.
\end{equation}
These implications cover every valuation $\nu_p(d)<2f$ in
all dimensions. For $\nu_p(d)\ge2f$, a three-dimensional
constant-weight section, endpoint lifting, and a surjective
functional produce a projective pencil whose known divisibilities
determine the remaining weight. Residual induction then decreases
the dimension and replaces $d$ by $d/q$. Establishing the basic
layers first in every dimension provides the starting cases for
this final induction.

\subsection{Organization}

Section~\ref{sec:coefficients} develops the reduced polynomials,
digit sums, line sums, and homogeneous coefficient forms used in
Section~\ref{sec:hyperplanes} to prove the hyperplane bound.
Section~\ref{sec:codes} proves the residual-code, endpoint, and
finite-difference lemmas needed for the coding argument.
Section~\ref{sec:window} establishes the basic divisibility
layers in all dimensions, and Section~\ref{sec:transfer}
extends them to all valuations, completing Theorem~\ref{thm:ward}.

\section{Prime-field degree and coefficient forms}\label{sec:coefficients}

The proof of the hyperplane bound uses two polynomial representations of
the same function. Field coordinates make scalar invariance visible,
whereas prime-field coordinates measure algebraic degree. We begin by
making the passage between these representations explicit.

\subsection{Reduced polynomials and the digit-degree filtration}

A polynomial in $\F_q[X_1,\ldots,X_m]$ is \emph{reduced} if its degree
in each individual variable is at most $q-1$. A polynomial and its
associated function must be distinguished: reduction preserves the
function, but may change its formal degree.

\begin{lemma}\label{lem:reduced-polynomial}
Every function $G:\F_q^m\to\F_q$ has exactly one reduced polynomial
representation. In particular, a reduced polynomial which vanishes at
every point of $\F_q^m$ is the zero polynomial.
\end{lemma}

\begin{proof}
For $a\in\F_q$, define
\[
 \delta_a(T)=1-(T-a)^{q-1}.
\]
Fermat's identity in $\F_q$ gives
\[
 \delta_a(t)=
 \begin{cases}
  1,&t=a,\\
  0,&t\ne a.
 \end{cases}
\]
Consequently,
\begin{equation}\label{eq:multivariate-interpolation}
 G(X_1,\ldots,X_m)
 =\sum_{a\in\F_q^m}G(a)\prod_{h=1}^m\delta_{a_h}(X_h)
\end{equation}
is a reduced polynomial representing $G$.

For uniqueness, a univariate polynomial of degree at most $q-1$ with
$q$ roots is zero. Induct on $m$ and write a reduced polynomial which
vanishes everywhere as
\[
 Q(X_1,\ldots,X_m)
 =\sum_{j=0}^{q-1}Q_j(X_1,\ldots,X_{m-1})X_m^j.
\]
For every $a\in\F_q^{m-1}$, the univariate polynomial $Q(a,X_m)$
vanishes at all $q$ elements of $\F_q$, so $Q_j(a)=0$ for every $j$.
Each $Q_j$ is reduced, and the induction hypothesis gives $Q_j=0$.
Thus $Q=0$, proving uniqueness.
\end{proof}

We now use this representation to compare the two coordinate descriptions
of a function. Fix an $\F_p$-basis $\beta_0,\ldots,\beta_{f-1}$ of $\F_q$ and
write
\[
 x_h=\sum_{j=0}^{f-1}\beta_j z_{h,j},
 \qquad z_{h,j}\in\F_p.
\]
Applying Lemma~\ref{lem:reduced-polynomial} over $\F_p$, with
coefficients allowed in $\F_q$, identifies the prime-field algebraic
degree with the total degree of the reduced polynomial in the
$fk$ variables $z_{h,j}$. An $\F_p$-linear substitution cannot
increase this degree: substitution preserves the degree bound, and
each reduction $Z^p=Z$ lowers degree. Therefore, for an invertible
$\F_p$-linear map $A$,
\[
 \deg_p(F\circ A)\le\deg_p F
 \quad\text{and}\quad
 \deg_p F=\deg_p\bigl((F\circ A)\circ A^{-1}\bigr)
 \le\deg_p(F\circ A).
\]
This proves basis independence and invariance under invertible
$\F_q$-linear coordinate changes. The same substitution argument
proves that restriction to an $\F_q$-linear subspace cannot increase
degree.

To use this degree without expanding every prime-field coordinate,
we express it directly in terms of the field exponents.
For a nonnegative integer $i$, let $\wt_p(i)$ denote its base-$p$
digit sum. The next identity is the standard $p$-weight description
of algebraic degree; compare the functional-degree formulation in
\cite[Theorem~10.3, journal version]{AichingerMoosbauer}. We retain a proof in terms
of reduced polynomial spaces.

\begin{lemma}\label{lem:digit-degree}
If the reduced polynomial of a nonzero function is
\[
 F(x_1,\ldots,x_k)=\sum_{\boldsymbol{i}}c_{\boldsymbol{i}}
 x_1^{i_1}\cdots x_k^{i_k},
\]
then
\[
 \deg_p F=\max_{c_{\boldsymbol{i}}\ne0}
       \sum_{h=1}^k\wt_p(i_h).
\]
Here, for an $\F_q$-valued function, degree is computed using
$\F_q$ coefficients in the reduced prime-field-coordinate polynomial.
If $F$ is scalar invariant, the ordinary total degree of every
nonconstant monomial with nonzero coefficient is divisible by $q-1$.
\end{lemma}

\begin{proof}
For $0\le i_h\le q-1$, write
\[
 i_h=\sum_{j=0}^{f-1}i_{h,j}p^j,
 \qquad 0\le i_{h,j}\le p-1.
\]
As functions of the prime-field coordinates, the Frobenius powers are $\F_p$-linear:
\[
 x_h^{p^j}
 =\left(\sum_{\ell=0}^{f-1}\beta_\ell z_{h,\ell}\right)^{p^j}
 =\sum_{\ell=0}^{f-1}\beta_\ell^{p^j}z_{h,\ell}.
\]
It follows that
\[
 x_1^{i_1}\cdots x_k^{i_k}
 =\prod_{h=1}^k\prod_{j=0}^{f-1}
      \bigl(x_h^{p^j}\bigr)^{i_{h,j}}
\]
has prime-field degree at most
$\sum_{h,j}i_{h,j}=\sum_h\wt_p(i_h)$.

To show that cancellation cannot lower the degree below the claimed
maximum, fix an integer $D\ge0$ and consider the following spaces of
$\F_q$-valued functions:
\begin{align*}
 \mathcal U_D
 &=\Span_{\F_q}\left\{
     x_1^{i_1}\cdots x_k^{i_k}:
     0\le i_h\le q-1,\
     \sum_h\wt_p(i_h)\le D\right\},\\
 \mathcal V_D
 &=\Span_{\F_q}\left\{
     \prod_{h,j}z_{h,j}^{\alpha_{h,j}}:
     0\le\alpha_{h,j}\le p-1,\
     \sum_{h,j}\alpha_{h,j}\le D\right\}.
\end{align*}
The preceding calculation gives $\mathcal U_D\subseteq\mathcal V_D$.
By uniqueness of reduced polynomials, each displayed spanning set is
linearly independent. Digit expansion provides the bijection
\[
 (i_1,\ldots,i_k)
 \longleftrightarrow(i_{h,j})_{h,j},
 \qquad
 \sum_h\wt_p(i_h)=\sum_{h,j}i_{h,j}.
\]
Hence $\dim_{\F_q}\mathcal U_D=\dim_{\F_q}\mathcal V_D$, and
$\mathcal U_D=\mathcal V_D$. Thus $\deg_p F\le D$ holds exactly
when every field monomial occurring in $F$ has digit sum at most $D$.
This proves the degree identity.

Finally, if $F(\lambda x)=F(x)$ for every $\lambda\in\F_q^*$,
both sides are reduced polynomials. Coefficient comparison gives
\[
 c_{\boldsymbol i}
 \bigl(\lambda^{i_1+\cdots+i_k}-1\bigr)=0
 \qquad(\lambda\in\F_q^*).
\]
For $c_{\boldsymbol i}\ne0$, cyclicity of $\F_q^*$ implies
$q-1\mid i_1+\cdots+i_k$, as required.
\end{proof}

\subsection{A truncated digit-sum inequality}

Scalar invariance restricts ordinary degrees to arithmetic
progressions with difference $q-1$. We next show that moving forward
in such a progression has a strictly positive digit cost. Keeping
arbitrarily many summands is essential for a bound independent of $k$.

\begin{lemma}\label{lem:cost}
For $n\ge0$, put
\[
 W_f(n)=p\left\lfloor\frac{n}{q}\right\rfloor
          +\wt_p(n\bmod q).
\]
If $0\le i_1,\ldots,i_r\le q-1$, then
\begin{equation}\label{eq:cost-bound}
 \sum_{h=1}^r\wt_p(i_h)\ge
 W_f\left(\sum_{h=1}^r i_h\right).
\end{equation}
Moreover,
\begin{equation}\label{eq:cost-monotone}
 W_f(n+q-1)>W_f(n)\qquad(n\ge0).
\end{equation}
\end{lemma}

\begin{proof}
Use the denominations $1,p,\ldots,p^{f-1}$, allowing an arbitrary
number of copies of each. Suppose a representation of $n$ contains
$p$ copies of $p^j$ with $j<f-1$. Replacing them by one copy of
$p^{j+1}$ preserves their sum and decreases the number of copies by
$p-1$. Repeating these operations terminates with a representation
\[
 n=c_{f-1}p^{f-1}+\sum_{j=0}^{f-2}c_jp^j,
 \quad c_{f-1}\ge0,\quad 0\le c_j\le p-1\ (j<f-1).
\]
This representation is unique by quotient and remainder on division
by $p^{f-1}$. Its number of summands is therefore minimal and equals
\begin{align*}
 c_{f-1}+\sum_{j=0}^{f-2}c_j
 &=\left\lfloor\frac{n}{p^{f-1}}\right\rfloor
     +\wt_p(n\bmod p^{f-1})\\
 &=p\left\lfloor\frac{n}{q}\right\rfloor
     +\wt_p(n\bmod q)=W_f(n).
\end{align*}
When $f=1$, the only denomination is $1$ and both expressions equal
$n$, so this case is included.

The base-$p$ expansions of $i_1,\ldots,i_r$ yield a representation
of $\sum_h i_h$ using exactly $\sum_h\wt_p(i_h)$ of these
denominations. Comparing with the minimum proves
\eqref{eq:cost-bound}.

For strict increase, write $n=uq+r$ with $0\le r<q$. If $r=0$,
then
\[
 W_f(n+q-1)-W_f(n)=\wt_p(q-1)=f(p-1)>0.
\]
If $r>0$, let $v=\nu_p(r)$. Subtracting $1$ replaces the $v$
trailing zero digits of $r$ by $p-1$ and decreases its next digit by
$1$. Hence
\[
 \wt_p(r-1)=\wt_p(r)-1+v(p-1).
\]
Since $n+q-1=(u+1)q+(r-1)$,
\begin{align*}
 W_f(n+q-1)-W_f(n)
 &=p+\wt_p(r-1)-\wt_p(r)\\
 &=(p-1)(v+1)>0.
\end{align*}
Thus $W_f$ is strictly increasing along every progression with step
$q-1$; monotonicity at consecutive integers is neither asserted nor
needed.
\end{proof}

\subsection{Line sums and formal linear factors}

The digit estimate will control polynomial coefficients. To choose the
coefficients that matter, we first examine the support on projective
lines.

For a scalar-invariant function $F:\F_q^k\to\F_p$ with $F(0)=0$,
define its \emph{projective support} by
\[
 S_F=\{\langle v\rangle\in\PG(k-1,q):F(v)\ne0\}.
\]
Scalar invariance makes $F(P):=F(v)$ well defined for
$P=\langle v\rangle$. We call a nonzero value $F(P)$ its
\emph{weight}; it lies in $\F_p^*$ and is distinct from an integer
Hamming weight. The following line identity is the elementary
prime-field instance of the low-degree summation principle underlying
\cite[Lemma~12.1, journal version]{AichingerMoosbauer}.

\begin{lemma}\label{lem:line-sum}
Under the hypotheses of Theorem~\ref{thm:hyperplane}, the values of
$F$ on every projective line sum to zero in $\F_p$. In particular,
no projective line meets $S_F$ in exactly one point.
\end{lemma}

\begin{proof}
Let $U\le\F_q^k$ have dimension $2$ over $\F_q$ and let
$\ell=\PG(U)$. In prime-field coordinates, $U\cong\F_p^{2f}$,
and the reduced polynomial of $F|_U$ has degree at most
\[
 (f+s)(p-1)\le(2f-1)(p-1)<2f(p-1).
\]
For $0\le j\le p-1$, the one-variable sum is
\begin{equation}\label{eq:prime-power-sums}
 \sum_{z\in\F_p}z^j=
 \begin{cases}
  0,&0\le j<p-1,\\
  -1,&j=p-1.
 \end{cases}
\end{equation}
For $j=0$ we use $z^0=1$, including $z=0$, and the sum is
$p=0$ in $\F_p$. For $1\le j<p-1$, the nonzero terms sum to
zero over the cyclic group $\F_p^*$; for $j=p-1$, they sum to
$p-1=-1$. This also proves the formula when $p=2$.

Summing a reduced monomial $Z_1^{\alpha_1}\cdots Z_{2f}^{\alpha_{2f}}$
over $\F_p^{2f}$ gives
\[
 \prod_{i=1}^{2f}\left(\sum_{z\in\F_p}z^{\alpha_i}\right).
\]
This product can be nonzero only if every $\alpha_i=p-1$, which
would give degree $2f(p-1)$. That degree exceeds the bound for
$F|_U$. Therefore
\[
 0=\sum_{v\in U}F(v)
   =F(0)+(q-1)\sum_{P\in\ell}F(P)
   =-\sum_{P\in\ell}F(P).
\]
The last equality uses $F(0)=0$ and $q-1=-1$ in $\F_p$.
If $\ell\cap S_F=\{P\}$, the last sum would equal the nonzero
element $F(P)$, a contradiction.
\end{proof}

Line sums will locate useful moments. Before extracting them, we
isolate the algebraic fact that turns functional vanishing into
formal divisibility. Its degree hypothesis prevents finite-field
polynomial identities from producing spurious factors.

\begin{lemma}\label{lem:linear-factors}
Let $Q\in\F_q[X_1,\ldots,X_m]$ have ordinary total degree less
than $q$, and let $L$ be a nonzero homogeneous linear polynomial.
If $Q(x)=0$ for every $x\in\ker L$, then $L$ divides $Q$ in
$\F_q[X_1,\ldots,X_m]$. Consequently, if $Q\ne0$, the number
of distinct linear hyperplanes on which $Q$ vanishes is at most
$\deg Q$.
\end{lemma}

\begin{proof}
Choose $A\in\operatorname{GL}_m(\F_q)$ with
$L(AZ)=Z_m$. The polynomial
\[
 Q(A(Z_1,\ldots,Z_{m-1},0))
\]
has total degree less than $q$, so each of its individual degrees
is less than $q$. It is therefore reduced and vanishes on
$\F_q^{m-1}$. By Lemma~\ref{lem:reduced-polynomial} it is the
zero polynomial. For $m=1$, this assertion simply says $Q(0)=0$.
Thus the coefficient of $Z_m^0$ in $Q(AZ)$ is zero, whence
$Z_m\mid Q(AZ)$ and $L\mid Q$.

Distinct hyperplanes have nonassociated linear defining equations.
These polynomials are irreducible and pairwise relatively prime in
the unique factorization domain $\F_q[X_1,\ldots,X_m]$. Their
product divides $Q$, so their number is at most $\deg Q$.
\end{proof}

\subsection{A coefficient polynomial for a pencil of hyperplanes}

The preceding lemma reduces hyperplane counting to finding a suitable
polynomial. We now construct that polynomial from a nonzero moment
on a line. The digit cost forces its ordinary degree to be strictly
less than $q$, so Lemma~\ref{lem:linear-factors} applies.

\begin{lemma}\label{lem:pencil}
Assume the hypotheses of Theorem~\ref{thm:hyperplane}, and put
$a=p^s$. Let $P\notin S_F$ lie on a projective line meeting $S_F$.
Take $P$ as the point at infinity on that line, let
$A\subseteq\F_q$ be its finite support coordinates, and let
$w_u\in\F_p^*$ be their weights. If
\[
 M_j=\sum_{u\in A}w_u u^j\ne0,
 \qquad j\ge1,\qquad p\nmid j,\qquad aj\le q-2,
\]
then at most $aj$ whole-zero hyperplanes of $F$ contain $P$.
\end{lemma}

\begin{proof}
Choose independent vectors $v_0,v_\infty$ representing the chosen
line so that
\[
 P=\langle v_\infty\rangle,
 \qquad u\longleftrightarrow\langle v_0+uv_\infty\rangle.
\]
Extend them to a basis of $\F_q^k$. In the resulting coordinates
$(x_1,\ldots,x_{k-1},y)$, the vector $v_\infty$ is the
$y$-direction and the finite line representatives have the form
$(x_0,u)$ for a fixed nonzero $x_0\in\F_q^{k-1}$. This
coordinate change preserves all the hypotheses. Write
\[
 F(x,y)=\sum_{b=0}^{q-1}C_b(x)y^b
\]
as its reduced polynomial, and put
\[
 e=aj=p^sj,\qquad b=q-1-e.
\]
The moment hypothesis gives $1\le e,b\le q-2$.

\emph{Step 1: The coefficient has ordinary degree $e$.}
Suppose $x_1^{i_1}\cdots x_{k-1}^{i_{k-1}}$ occurs in $C_b$.
Lemma~\ref{lem:digit-degree} gives
\begin{align}
 \sum_{h=1}^{k-1}\wt_p(i_h)
 &\le(f+s)(p-1)-\wt_p(b)\notag\\
 &=s(p-1)+\wt_p(e)
  =s(p-1)+\wt_p(j)=:B.\label{eq:coefficient-budget}
\end{align}
Here the equalities follow from
\[
 \wt_p(q-1-e)=f(p-1)-\wt_p(e),
 \qquad \wt_p(p^sj)=\wt_p(j),
\]
because $e<q$ and multiplication by $p^s$ shifts the digits of
$j$ without carrying. Scalar invariance also gives
\[
 \sum_{h=1}^{k-1}i_h+b\equiv0\pmod{q-1},
 \quad\text{hence}\quad
 \sum_{h=1}^{k-1}i_h=e+t(q-1)
 \quad(t\ge0).
\]
The integer $t$ is nonnegative since $0<e<q-1$ and the left
side is nonnegative.

The condition $p\nmid j$ makes its lowest digit nonzero. Thus
\[
 e-1=(j-1)p^s+(p^s-1),
 \qquad
 \wt_p(e-1)=\wt_p(j)-1+s(p-1)=B-1.
\]
Since $e+q-1=q+(e-1)$, we obtain
\begin{equation}\label{eq:exclude-next-degree}
 W_f(e+q-1)=p+\wt_p(e-1)=B+(p-1)>B.
\end{equation}
If $t\ge1$, Lemma~\ref{lem:cost} would imply
\[
 \sum_h\wt_p(i_h)
 \ge W_f\bigl(e+t(q-1)\bigr)
 \ge W_f(e+q-1)>B,
\]
contradicting \eqref{eq:coefficient-budget}. Therefore $t=0$.
Every monomial in $C_b$ has ordinary total degree exactly $e$;
in particular, $C_b$ is formally homogeneous of degree $e<q$
unless it is zero.

\emph{Step 2: The coefficient is nonzero.}
On the selected affine line, interpolation gives
\[
 F(x_0,y)=\sum_{u\in A}w_u\bigl(1-(y-u)^{q-1}\bigr).
\]
Both sides are reduced in $y$, so their coefficients agree. In
characteristic $p$,
\[
 (1+T)\sum_{r=0}^{q-1}(-T)^r
 =1-(-T)^q=1+T^q=(1+T)^q.
\]
Canceling $1+T$ and comparing coefficients gives
$\binom{q-1}{r}=(-1)^r$ in $\F_q$.
Since $b\ge1$ and $b+e=q-1$, comparison of the coefficient
of $y^b$ yields
\begin{align*}
 C_b(x_0)
 &=-\binom{q-1}{b}(-1)^e\sum_{u\in A}w_u u^e\\
 &=-\sum_{u\in A}w_u u^{p^sj}\\
 &=-\left(\sum_{u\in A}w_u u^j\right)^{p^s}
  =-M_j^a\ne0.
\end{align*}
We used $(-1)^{q-1}=1$ in $\F_q$ and $w_u^{p^s}=w_u$,
the latter following from $w_u\in\F_p$.

\emph{Step 3: Each whole-zero hyperplane contributes a linear factor.}
A hyperplane containing the $y$-direction has equation $L(x)=0$
for a nonzero homogeneous linear form in the $x$ variables. If
it is a whole-zero hyperplane, then, for every $x\in\ker L$,
the reduced univariate polynomial
\[
 \sum_{r=0}^{q-1}C_r(x)y^r
\]
vanishes for all $y\in\F_q$. Uniqueness gives $C_r(x)=0$ for
every $r$, including $r=b$. Thus $C_b$ vanishes as a function
on $\ker L$. Its ordinary degree is $e<q$, so
Lemma~\ref{lem:linear-factors} gives $L\mid C_b$ as a formal
polynomial. Since $C_b\ne0$ and $\deg C_b=e$, at most $e=aj$
distinct whole-zero hyperplanes can contain $P$.
\end{proof}

\section{The dimension-free zero-hyperplane bound}\label{sec:hyperplanes}

With the coefficient lemma in place, it remains to pass from bounds
through one outside point to a bound for all whole-zero hyperplanes.
For a short support secant, this will require control of the moment
indices at all its outside points. If every support secant is too long,
the size of the support gives the desired estimate by a second moment.
We establish this incidence estimate first.

\subsection{A projective incidence estimate}

The next bound is purely geometric. Its proof uses a projective
version of the support and second-moment method in
\cite[Lemma~4.8 and Section~4.2.3]{HRS}, together with a sharper
Cauchy--Schwarz estimate. The precise numerical statement below
is proved directly.

\begin{lemma}\label{lem:incidence}
Let $S$ be a nonempty set of points in $\PG(k-1,q)$, where $k\ge2$.
If $N$ distinct projective hyperplanes are disjoint from $S$, then
\[
 N\le\frac{q^k}{|S|}-(q-1).
\]
\end{lemma}

\begin{proof}
If $N=0$, then $|S|\le(q^k-1)/(q-1)$ makes the right side
positive. Assume $N>0$, and choose a nonzero defining linear form
$L_i$ for each hyperplane, $1\le i\le N$. Let $X$ be uniform
on $\F_q^k$, and set
\[
 I_i=\mathbf 1_{\{L_i(X)=0\}},
 \qquad Y=\sum_{i=1}^N I_i.
\]
Every $L_i$ is surjective, and distinct projective hyperplanes
have linearly independent defining forms. Therefore
\[
 \E I_i=\frac1q,
 \qquad
 \E(I_iI_j)=\frac1{q^2}\quad(i\ne j).
\]
Using $I_i^2=I_i$ gives
\begin{align*}
 \E Y&=\frac Nq,\\
 \E Y^2
 &=\sum_i\E I_i+\sum_{i\ne j}\E(I_iI_j)\\
 &=\frac Nq+\frac{N(N-1)}{q^2}
  =\frac{N(N+q-1)}{q^2}.
\end{align*}
Cauchy--Schwarz applied to $Y\mathbf 1_{\{Y>0\}}$ gives
\[
 (\E Y)^2\le\Prob(Y>0)\E Y^2,
 \qquad
 \Prob(Y=0)\le1-\frac{N}{N+q-1}
               =\frac{q-1}{N+q-1}.
\]
The set of nonzero vectors representing points of $S$ has
cardinality $(q-1)|S|$. Every such vector lies in the event $Y=0$,
because each chosen hyperplane is disjoint from $S$. Hence
\[
 \frac{(q-1)|S|}{q^k}
 \le\Prob(Y=0)\le\frac{q-1}{N+q-1}.
\]
Canceling $q-1$ and rearranging proves the estimate.
\end{proof}

\subsection{A common budget for the moment indices}

The incidence estimate will handle the case of large support. We now
return to a single projective line and control the indices needed to
sum the bounds from Lemma~\ref{lem:pencil}. The rational function
below records all positive-index moments in its expansions at the
possible choices of infinity.

\begin{lemma}\label{lem:moment-budget}
Let $A\subseteq\F_q$ have size $n\ge2$, and let
$w_u\in\F_p^*$ satisfy $\sum_{u\in A}w_u=0$.
For the point at infinity, use support coordinates $u$; for a
finite point $t\in\F_q\setminus A$, use support coordinates
$(u-t)^{-1}$. For each outside point
$P\in\PG(1,q)\setminus A$, the first nonzero positive-index
moment exists. If its index is $j_P$, then
\[
 p\nmid j_P,\qquad 1\le j_P\le n-1,
 \qquad
 \sum_{P\in\PG(1,q)\setminus A}(j_P-1)\le n-2.
\]
\end{lemma}

\begin{proof}
Define
\begin{equation}\label{eq:rational-moments}
 h(T)=\prod_{u\in A}(T-u),\qquad
 R(T)=\sum_{u\in A}\frac{w_u}{T-u}=\frac{B(T)}{h(T)},
\end{equation}
where explicitly
\[
 B(T)=\sum_{u\in A}w_u
              \prod_{v\in A\setminus\{u\}}(T-v).
\]
For each $u\in A$,
\[
 B(u)=w_u\prod_{v\in A\setminus\{u\}}(u-v)\ne0.
\]
Thus $B\ne0$ and no factor $T-u$ cancels. The coefficient of
$T^{n-1}$ is $\sum_u w_u=0$, so
\[
 0\le\deg B\le n-2.
\]

For a finite outside point $t$, let
\[
 r_t=\ord_{T=t}B(T)
 =\max\{r\ge0:(T-t)^r\mid B(T)\}.
\]
Since $h(t)\ne0$, this is also
the order of $R$ at $t$. Expanding geometrically in the formal
power-series ring $\F_q[[U]]$ gives
\begin{align}
 R(t+U)
 &=\sum_{u\in A}\frac{-w_u}{u-t}
       \frac1{1-U/(u-t)}\notag\\
 &=-\sum_{m\ge0}
       \left(\sum_{u\in A}w_u(u-t)^{-m-1}\right)U^m.
       \label{eq:finite-moment-expansion}
\end{align}
Consequently the moments of indices $1,\ldots,r_t$ vanish and
the moment of index $r_t+1$ is nonzero. Thus $j_t=r_t+1$.

At the original point at infinity,
\[
 R(1/U)=\sum_{u\in A}\frac{w_uU}{1-uU}
        =\sum_{m\ge0}\left(\sum_{u\in A}w_u u^m\right)U^{m+1}.
\]
The term with $m=0$ vanishes. Dividing by $U^2$ therefore yields
\begin{equation}\label{eq:infinite-moment-expansion}
 \frac{R(1/U)}{U^2}
 =\sum_{m\ge0}\left(\sum_{u\in A}w_u u^{m+1}\right)U^m.
\end{equation}
Since $h$ is monic of degree $n$ and $B\ne0$, the left side
has exact order
\[
 r_\infty=n-2-\deg B\ge0.
\]
Hence $j_\infty=r_\infty+1$. These calculations also prove
existence of a first nonzero positive-index moment at every
outside point.

In any of the displayed coordinate systems, prime-field weights
give the Frobenius relation
\[
 M_{pv}=\sum_u w_u z_u^{pv}
       =\left(\sum_u w_u z_u^v\right)^p=M_v^p.
\]
If the first nonzero positive-index moment had index $pv$, then
$M_v\ne0$ would contradict its minimality. Therefore $p\nmid j_P$
for every $P$.

Finally, distinct finite outside points contribute distinct linear
factors to $B$, so
\[
 \sum_{t\in\F_q\setminus A}r_t\le\deg B.
\]
Including infinity gives the common budget
\begin{equation}\label{eq:root-budget}
 \sum_{P\in\PG(1,q)\setminus A}(j_P-1)
 =\sum_{t\in\F_q\setminus A}r_t+r_\infty
 \le\deg B+(n-2-\deg B)=n-2.
\end{equation}
All summands are nonnegative, so each $j_P\le n-1$ as well.
\end{proof}

\subsection{Proof of the hyperplane bound}

The moment budget is the remaining ingredient for summing the pencil
bounds on a short support secant. When no such secant exists, the
incidence estimate applies instead. We now carry out this split for
the function in Theorem~\ref{thm:hyperplane}.

\begin{proof}[Proof of Theorem~\ref{thm:hyperplane}]
Let $S=S_F$, let $N$ be the number of whole-zero hyperplanes,
and put $a=p^s$. Since $F\ne0$ and $F(0)=0$, the support
$S$ is nonempty.

For $k=2$, the entire projective space is a line. By
Lemma~\ref{lem:line-sum}, $|S|\ge2$. Hyperplanes are then
projective points, and therefore
\[
 N=(q+1)-|S|\le q-1\le a(q-1).
\]
Assume henceforth that $k\ge3$.

\medskip\noindent
\emph{Case 1: a short support secant exists.}
Suppose a projective line $\ell$ has $n$ support points, with
\begin{equation}\label{eq:short-secant-range}
 \begin{cases}
  2\le n\le q/a,&a\ge2,\\
  2\le n\le q-1,&a=1.
 \end{cases}
\end{equation}
In either case $n<q+1$, so $\ell$ has an outside point. Choose
one as infinity and identify its finite support with a set
$A\subseteq\F_q$ of size $n$. Let $w_u=F(P_u)\in\F_p^*$
be the support values. Lemma~\ref{lem:line-sum} gives
$\sum_{u\in A}w_u=0$, so Lemma~\ref{lem:moment-budget}
provides indices $j_P$ at all $q+1-n$ outside points.

These indices satisfy $p\nmid j_P$, and their ranges permit
the use of Lemma~\ref{lem:pencil}:
\[
 aj_P\le a(n-1)\le q-a\le q-2\quad(a\ge2),
\]
whereas for $a=1$,
\[
 j_P\le n-1\le q-2.
\]
Each projective coordinate change on $\ell$ comes from an
invertible linear transformation of its two-dimensional vector
space, which extends to an invertible linear transformation of
$\F_q^k$. Scalar invariance preserves the weights when
representatives are rescaled, and prime-field degree is preserved
under the ambient transformation. Explicitly, if the original representatives are
$v_0+uv_\infty$ and the new point at infinity has finite coordinate
$t\notin A$, choose
\[
 v'_0=v_\infty,\qquad v'_\infty=v_0+tv_\infty.
\]
Then, for every $u\in A$,
\[
 v_0+uv_\infty
 =(u-t)\bigl(v'_0+(u-t)^{-1}v'_\infty\bigr).
\]
The two new vectors are independent, and scalar invariance removes
the prefactor $u-t$. This realizes precisely the finite-point
coordinates in Lemma~\ref{lem:moment-budget}.
Thus the coefficient lemma applies separately at every outside
point $P$ and gives
\[
 \#\{H:F|_H\equiv0,\ P\in\PG(H)\}\le aj_P.
\]

A whole-zero hyperplane cannot contain $\ell$, because $\ell$
contains support points. Since $k\ge3$, it meets $\ell$ at
exactly one point, necessarily outside $S$. Therefore every
whole-zero hyperplane is counted exactly once when its intersection
with $\ell$ is used, and
\begin{align*}
 N
 &\le a\sum_{P\in\ell\setminus S}j_P\\
 &=a\left((q+1-n)+\sum_{P\in\ell\setminus S}(j_P-1)\right)\\
 &\le a\bigl((q+1-n)+(n-2)\bigr)=a(q-1).
\end{align*}

\medskip\noindent
\emph{Case 2: no secant in \eqref{eq:short-secant-range} exists.}
Fix $P_0\in S$. The projective lines through $P_0$ correspond
to points of the quotient projective space $\PG(k-2,q)$, so
their number is
\[
 \theta_{k-2}=\frac{q^{k-1}-1}{q-1}.
\]
Two different such lines meet only in $P_0$, and every other
projective point lies on exactly one of them. Every one of these
lines has at least two support points by
Lemma~\ref{lem:line-sum}.

If $a=1$, the absence of intersections of sizes $2,\ldots,q-1$
means that every line through $P_0$ has at least $q$ support
points. Consequently
\[
 |S|\ge1+(q-1)\theta_{k-2}=q^{k-1}.
\]
Every whole-zero hyperplane is disjoint from $S$, so
Lemma~\ref{lem:incidence} gives
\[
 N\le\frac{q^k}{|S|}-(q-1)
   \le q-(q-1)=1\le q-1.
\]
This includes $q=2$: the short-secant range is empty, and the
line-sum identity still ensures at least $q=2$ support points
on every line through $P_0$.

If $a\ge2$, then $q/a=p^{f-s}$ is an integer, and every line
through $P_0$ has at least $q/a+1$ support points. Thus
\begin{align*}
 |S|
 &\ge1+\frac qa\theta_{k-2}\\
 &=\frac{q^k+(a-1)q-a}{a(q-1)}.
\end{align*}
Since $s<f$, we have $q\ge pa\ge2a$; hence
\[
 (a-1)q-a\ge2a(a-1)-a=a(2a-3)>0.
\]
It follows that
\[
 \frac{q^k}{|S|}<a(q-1).
\]
Applying Lemma~\ref{lem:incidence} once more gives
\[
 N\le\frac{q^k}{|S|}-(q-1)<a(q-1).
\]
The two cases exhaust all support configurations, completing the
proof.
\end{proof}

The first-layer value of the bound is attained in every dimension,
as the following example shows.

\begin{remark}\label{rem:sharp-first}
For $s=0$ and $k\ge2$, take
\[
 F(x_1,\ldots,x_k)=x_1^{q-1}-x_2^{q-1}.
\]
Since $t^{q-1}\in\{0,1\}$ for $t\in\F_q$, the function
has values in $\F_p$. It satisfies
\[
 F(0)=0,\qquad F(1,0,\ldots,0)=1,
 \qquad F(\lambda x)=F(x)\quad(\lambda\ne0).
\]
Lemma~\ref{lem:digit-degree} gives
$\deg_p F=\wt_p(q-1)=f(p-1)$. On the hyperplane
$x_1=\lambda x_2$, with $\lambda\in\F_q^*$,
\[
 F(x)=(\lambda^{q-1}-1)x_2^{q-1}=0.
\]
The $q-1$ defining equations have coefficient $1$ at $x_1$
and distinct coefficients at $x_2$, so they define distinct
hyperplanes. Theorem~\ref{thm:hyperplane} gives the matching
upper bound $q-1$.
\end{remark}

\section{Griesmer codes, endpoints, and weight residues}\label{sec:codes}

To apply the geometric theorem to code divisibility, we pass from
integer weights to functions on the coefficient space. Residual codes
and endpoint shortenings provide the smaller Griesmer codes needed
for induction; divided weight residues supply the prime-field
polynomials to which the hyperplane bound will apply. The following
arguments develop these two ingredients and establish the required
degree bound.
The residual and endpoint statements are classical; see
\cite[Section 4]{Ward2001} and
\cite[arXiv version, Lemma 3.1, Theorem 3.3, and Corollary 3.4]{DHX}.
Their proofs are retained, including the dimension and effective-length
arguments needed later.

\subsection{Residual codes and the Griesmer bound}

If $I\subseteq\{1,\ldots,n\}$, puncturing the coordinates in
$I$ means applying the linear map
\[
 \pi_I:\F_q^n\longrightarrow\F_q^{\{1,\ldots,n\}\setminus I},
 \qquad (c_i)_{i=1}^n\longmapsto(c_i)_{i\notin I}.
\]
When $I=\supp(b)$ for a minimum-weight word $b$, its image on
$C$ is called the residual code with respect to $b$.
The minimum-weight hypothesis controls both the kernel of this
puncturing map and the distance of its image, leading directly to
the Griesmer bound.

\begin{lemma}\label{lem:residual}
Let $C$ be an $[n,k,d]_q$ linear code with $k\ge1$ and $d\ge1$.
Then $n\ge g_q(k,d)$. If $k\ge2$ and $b\in C$ has weight $d$,
puncturing the positions in $\supp(b)$ gives a code of dimension
$k-1$ and minimum distance at least $\lceil d/q\rceil$.
If $C$ is Griesmer, this residual code is Griesmer of minimum
distance exactly $\lceil d/q\rceil$; in particular its distance
is $d/q$ when $q\mid d$.
\end{lemma}

\begin{proof}
In dimension one the Griesmer bound is $n\ge d$, and no residual
assertion is made. For the residual assertions assume $k\ge2$.
Put $I=\supp(b)$ and
$C_b=\pi_I(C)$. For $u\in C\setminus\Span(b)$, set
\[
 r=\wt(\pi_I(u))=|\{i\notin I:u_i\ne0\}|.
\]
Every word $u+\alpha b$, with $\alpha\in\F_q$, is nonzero;
otherwise $u=-\alpha b\in\Span(b)$. For $i\in I$, the
equation $u_i+\alpha b_i=0$ has the unique solution
$\alpha=-u_i/b_i$. For $i\notin I$, its left side equals
$u_i$ independently of $\alpha$. Thus the coordinate sums are
\[
 \sum_{\alpha\in\F_q}{\bf1}_{\{u_i+\alpha b_i\ne0\}}
 =\begin{cases}
 q-1,&i\in I,\\
 q,&i\notin I\text{ and }u_i\ne0,\\
 0,&i\notin I\text{ and }u_i=0.
 \end{cases}
\]
Summing over the coordinates gives the following identity. It
provides a direct proof of the residual bound stated in
\cite[arXiv v1, Lemma 3.1]{DHX}:
\begin{equation}\label{eq:residual-average}
 \sum_{\alpha\in\F_q}\wt(u+\alpha b)
 =(q-1)d+qr\ge qd.
\end{equation}
Consequently,
\[
 qr\ge d,\qquad
 r\ge\left\lceil\frac dq\right\rceil\ge1.
\]

The kernel and the dimension now follow without any assumption on
the multiplicities of the generator columns. Certainly
$\Span(b)\subseteq\ker(\pi_I|_C)$. Conversely, a word
$u\notin\Span(b)$ has $\wt(\pi_I(u))\ge1$, so it cannot
belong to that kernel. Hence
\[
 \ker(\pi_I|_C)=\Span(b),\qquad
 \dim C_b=\dim C-\dim\Span(b)=k-1.
\]
Every nonzero word of $C_b$ has a preimage outside $\Span(b)$;
the same inequality therefore proves
$d(C_b)\ge\lceil d/q\rceil$.

We next obtain the Griesmer bound by induction on $k$.
For $k=1$, the inequality is $n\ge d=g_q(1,d)$.
If the bound holds in dimension $k-1$, then
\begin{align*}
 n-d
 &\ge g_q(k-1,d(C_b))\\
 &\ge g_q\left(k-1,\left\lceil\frac dq\right\rceil\right)
 =\sum_{i=0}^{k-2}
      \left\lceil\frac{\lceil d/q\rceil}{q^i}\right\rceil
 =\sum_{i=1}^{k-1}\left\lceil\frac{d}{q^i}\right\rceil.
\end{align*}
Here we used the elementary identity
\[
 \left\lceil\frac{\lceil d/q\rceil}{q^i}\right\rceil
 =\left\lceil\frac{d}{q^{i+1}}\right\rceil,
 \qquad i\ge0.
\]
Adding $d$ proves $n\ge g_q(k,d)$.

Finally, suppose that $n=g_q(k,d)$ and put $d_0=\lceil d/q\rceil$.
The ceiling identity just proved gives
\[
 n-d=g_q(k-1,d_0).
\]
For every positive integer $m$, the function
$x\mapsto g_q(m,x)$ is strictly increasing on positive integers:
its first summand is $x$, and its remaining summands are
nondecreasing. If $d(C_b)>d_0$, the Griesmer bound just proved
would give
\[
 n-d\ge g_q(k-1,d(C_b))>g_q(k-1,d_0)=n-d,
\]
a contradiction. Thus $d(C_b)=d_0$, and $C_b$ is Griesmer.
When $q\mid d$, this gives $d(C_b)=d/q$.
This is \cite[arXiv v1, Theorem 3.3(1)]{DHX}.
\end{proof}

\subsection{Projective columns and endpoint shortenings}

A complementary construction reduces dimension by restricting the
coefficient space while preserving the minimum distance. Its
description uses the projective classes of the generator columns.

Fix a Griesmer code $C$ and a full-rank generator matrix $G$.
Its coefficient space is $V=\F_q^k$, with the isomorphism
\[
 \iota_G:V\longrightarrow C,\qquad u\longmapsto uG.
\]
We shall write $\wt(u)=\wt(uG)$ and
$\supp(u)=\supp(uG)$. The $i$th column of $G$
represents the coordinate functional
\[
 \lambda_i\in V^*,\qquad \lambda_i(u)=(uG)_i.
\]
No $\lambda_i$ is zero. Indeed, deleting a zero column would
leave both dimension and minimum distance unchanged and produce
an $[g_q(k,d)-1,k,d]_q$ code, contrary to
Lemma~\ref{lem:residual}.

For a vector space $W$, the projective space $\PG(W)$ is the
set of one-dimensional subspaces of $W$. Thus a point of
$\PG(V^*)$ is a class
\[
 P=\langle\lambda\rangle=\F_q\lambda,
 \qquad\lambda\in V^*\setminus\{0\}.
\]
Let $M(P)$ be the number of columns whose functionals represent
$P$; repetitions are counted. For a set $S\subseteq\PG(V^*)$,
put $M(S)=\sum_{P\in S}M(P)$. In particular,
\[
 M(\PG(V^*))=n.
\]
For $u\in V\setminus\{0\}$, the annihilating projective
hyperplane is
\[
 K_u=\{\langle\lambda\rangle\in\PG(V^*):\lambda(u)=0\}.
\]
The coordinates counted by $M(K_u)$ are precisely the zero
coordinates of $uG$, and therefore
\begin{equation}\label{eq:weight-columns}
 \wt(u)=n-M(K_u),\qquad
 M(K_u)=\sum_{P\in K_u}M(P).
\end{equation}
Every projective hyperplane of $\PG(V^*)$ is $K_u$ for some
nonzero $u$, by the natural identification $(V^*)^*=V$.

For later restrictions, let $U\le V$ and define
\[
 I(U)=\{i:\lambda_i|_U=0\}.
\]
The \emph{effective code on $U$} is the image of
\[
 \iota_U:U\longrightarrow\F_q^{\{1,\ldots,n\}\setminus I(U)},
 \qquad u\longmapsto(\lambda_i(u))_{i\notin I(U)}.
\]
This map is injective. In fact, if $\iota_U(u)=0$, the retained
coordinates vanish by assumption and the deleted coordinates
vanish because $u\in U$; hence $uG=0$ and $u=0$.
Consequently,
\[
 \dim\iota_U(U)=\dim U,
 \qquad \wt(\iota_U(u))=\wt(uG)\quad(u\in U).
\]
Restoring the deleted zero coordinates identifies this effective
code with the subcode $\iota_G(U)\le C$. In particular, its
minimum distance is at least $d$ whenever $U\ne\{0\}$.
This is the effective-length convention used in
\cite[Section 1]{Ward2001}.

The points of maximum column multiplicity specify precisely the
codimension-one restrictions needed for induction. The next
statement includes Ward's endpoint count
\cite[Lemma 11]{Ward2001}, together with the standard shortening
and multiplicity facts
\cite[arXiv version, Theorem 3.3(2) and Corollary 3.4]{DHX}.

\begin{lemma}\label{lem:endpoints}
Let $C$ be a Griesmer code of dimension $k\ge2$. Put
\[
 Q=q^{k-1},\qquad
 \mu=\left\lceil\frac{d}{Q}\right\rceil,\qquad
 t=d-(\mu-1)Q\quad(1\le t\le Q).
\]
Then $M(P)\le\mu$ for every $P\in\PG(V^*)$. The
\emph{endpoint set}
\[
 E=\{P\in\PG(V^*):M(P)=\mu\}
\]
satisfies
\begin{equation}\label{eq:endpoint-bound}
 |E\setminus K|\ge t\quad\text{for every hyperplane }K,
 \qquad |E|\ge g_q(k,t)\ge t.
\end{equation}
In particular, $E$ is nonempty and spans $\PG(V^*)$.
For $P=\langle\lambda\rangle\in E$, the effective code on
$\ker\lambda$ is a dimension-$(k-1)$ Griesmer code of minimum
distance exactly $d$. If $Q\mid d$, every nonzero word of $C$
has weight $d$.
\end{lemma}

\begin{proof}
The definition of $\mu$ gives
\[
 (\mu-1)Q<d\le\mu Q,
 \qquad
 0<t=d-(\mu-1)Q\le Q.
\]
Since $t$ is an integer, this also proves the stated range
$1\le t\le Q$.

For an arbitrary $P=\langle\lambda\rangle$, take
$U=\ker\lambda$, a subspace of dimension $k-1$.
Its annihilator is
\[
 U^\perp=\{\gamma\in V^*: \gamma|_U=0\}
          =\Span(\lambda).
\]
Thus the identically zero coordinates on $U$ are exactly the
$M(P)$ columns represented by $P$. Its effective length is
$n-M(P)$, and its minimum distance $d_U$ satisfies $d_U\ge d$.
Applying Lemma~\ref{lem:residual} gives
\begin{align*}
 n-M(P)&\ge g_q(k-1,d_U)\ge g_q(k-1,d),\\
 M(P)&\le g_q(k,d)-g_q(k-1,d)
       =\left\lceil\frac{d}{q^{k-1}}\right\rceil=\mu.
\end{align*}
When $P\in E$, equality of the effective length with
$g_q(k-1,d)$ and strict increase of the Griesmer sum force
$d_U=d$. This proves the shortening assertion.

We next count the endpoints. A projective space of vector
dimension $k$ has $(q^k-1)/(q-1)$ points; a projective
hyperplane has $(q^{k-1}-1)/(q-1)$ points. Therefore
\[
 |\PG(V^*)\setminus K|=q^{k-1}=Q.
\]
Since every nonendpoint has integer multiplicity at most
$\mu-1$, equation~\eqref{eq:weight-columns} implies
\begin{align*}
 d
 &\le M(\PG(V^*)\setminus K)\\
 &\le \mu|E\setminus K|
       +(\mu-1)\bigl(Q-|E\setminus K|\bigr)\\
 &=(\mu-1)Q+|E\setminus K|.
\end{align*}
Subtracting $(\mu-1)Q$ gives $|E\setminus K|\ge t$.
In particular, no projective hyperplane contains $E$; this is
equivalent to the assertion that its representatives span $V^*$.

Counting all points, rather than only those outside $K$, gives
Ward's sharper cardinality estimate:
\begin{align*}
 n
 &=\sum_{P\in\PG(V^*)}M(P)\\
 &\le\mu|E|+(\mu-1)
       \left(\frac{q^k-1}{q-1}-|E|\right),\\
 |E|
 &\ge n-(\mu-1)\frac{q^k-1}{q-1}\\
 &=\sum_{i=0}^{k-1}
       \left(\left\lceil\frac{d}{q^i}\right\rceil
                  -(\mu-1)q^{k-1-i}\right)\\
 &=\sum_{i=0}^{k-1}
       \left\lceil\frac{d-(\mu-1)q^{k-1}}{q^i}\right\rceil
   =g_q(k,t).
\end{align*}
In the penultimate equality, the number
$(\mu-1)q^{k-1-i}$ is an integer, so it can be moved
inside the ceiling. The first summand of $g_q(k,t)$ is $t$,
proving the remaining inequality in
\eqref{eq:endpoint-bound}.

Finally, if $Q\mid d$, then $\mu=d/Q$. For every
$u\ne0$, the complement of $K_u$ has $Q$ points, each
of multiplicity at most $\mu$. Hence
\[
 d\le\wt(u)=M(\PG(V^*)\setminus K_u)
             \le\mu Q=d.
\]
All these inequalities are equalities, so $C$ is
constant-weight. The same statement appears in
\cite[arXiv v1, Proposition 4.1]{DHX}, where it is attributed
to \cite[Corollary 5]{Ward1998}.
\end{proof}

\subsection{Finite differences and weight residues}

Endpoint shortenings describe where a weight residue must vanish.
To apply Theorem~\ref{thm:hyperplane}, we also need a bound on
its prime-field degree. The relevant finite-difference estimate
is the elementary-abelian case of
\cite[Theorem 4.9(c)]{ClarkSchauz}. We first isolate it with
the coefficient ring made explicit.

For an abelian group $A$ and a function $h:A\to R$, write
\[
 (T_vh)(x)=h(x+v),\qquad
 (\delta_vh)(x)=h(x+v)-h(x),\qquad v,x\in A.
\]
The translation operators commute, and $\delta_v=T_v-I$.
An iterated difference has the explicit formula
\begin{equation}\label{eq:iterated-difference}
 (\delta_{v_1}\cdots\delta_{v_m}h)(x)
 =\sum_{J\subseteq\{1,\ldots,m\}}
       (-1)^{m-|J|}h\left(x+\sum_{j\in J}v_j\right).
\end{equation}
We shall use this formula both before and after dividing
integer weights by a power of $p$.

For a nonzero function $h$, its \emph{functional degree} is
\[
 \operatorname{fdeg}(h)
 =\min\{D\in\Z_{\ge0}:
       \delta_{v_1}\cdots\delta_{v_{D+1}}h=0
       \text{ for all }v_1,\ldots,v_{D+1}\in A\},
\]
with value $\infty$ if the set is empty. We set
$\operatorname{fdeg}(0)=-\infty$. The next lemma is an upper
bound on this degree for a specified finite domain and target.

\begin{lemma}\label{lem:additive-differences}
Let $s$ be a nonnegative integer, let $R=\Z/p^{s+1}\Z$, and let $A$ be the
additive group of $\F_q$, where $q=p^f$. Put
\[
 L=(f+s)(p-1)+1.
\]
For every function $h:A\to R$ and all $v_1,\ldots,v_L\in A$,
\[
 \delta_{v_1}\cdots\delta_{v_L}h=0.
\]
\end{lemma}

\begin{proof}
Choose an $\F_p$-basis $\varepsilon_1,\ldots,\varepsilon_f$
of $\F_q$. On the $R$-module of all functions $A\to R$,
put
\[
 T_i=T_{\varepsilon_i},\qquad D_i=T_i-I.
\]
The operators commute, and $p\varepsilon_i=0$ implies
$T_i^p=I$. Expanding $(I+D_i)^p=I$ gives
\begin{equation}\label{eq:difference-reduction}
 D_i^p=-\sum_{j=1}^{p-1}\binom pj D_i^j.
\end{equation}
For $1\le j\le p-1$, the integer $\binom pj$ is divisible
by $p$.

Consider an operator monomial
$D_1^{a_1}\cdots D_f^{a_f}$ with
$A_0=a_1+\cdots+a_f\ge L$. Whenever an exponent is at least
$p$, use \eqref{eq:difference-reduction} on one factor
$D_i^p$. Each resulting term gains a factor divisible by
$p$, and its total exponent decreases by one of
$1,\ldots,p-1$. Repeating this replacement terminates,
because total exponents strictly decrease. At the end of a
branch every exponent is at most $p-1$, so its final total
exponent $A_r$ satisfies $A_r\le f(p-1)$.
If that branch contains $r$ replacements, then
\[
 r(p-1)\ge A_0-A_r
          \ge L-f(p-1)=s(p-1)+1.
\]
It follows that $r\ge s+1$. The branch coefficient is
therefore divisible by $p^{s+1}$ and is zero in $R$.
Every branch vanishes, proving
\begin{equation}\label{eq:operator-monomial-vanishing}
 D_1^{a_1}\cdots D_f^{a_f}=0
 \quad\text{if}\quad \sum_{i=1}^f a_i\ge L.
\end{equation}

For an arbitrary $v\in A$, write
$v=\sum_{i=1}^f v_i\varepsilon_i$ with integer representatives
$0\le v_i<p$. Its translation difference is
\[
 \delta_v=T_v-I
  =\prod_{i=1}^f T_i^{v_i}-I
  =\prod_{i=1}^f(I+D_i)^{v_i}-I.
\]
This is a polynomial in $D_1,\ldots,D_f$ with zero constant
term. A product of $L$ such differences is consequently a
sum of operator monomials of total exponent at least $L$.
Equation~\eqref{eq:operator-monomial-vanishing} makes every
one of them zero. Applying the resulting zero operator to
$h$ proves the assertion.
\end{proof}

Writing $\delta(A,B)$ for the largest functional degree of maps
$A\to B$ in this finite $p$-group setting, the general formula of
Clark and Schauz gives
\[
 \delta\bigl((\Z/p\Z)^f,\Z/p^{s+1}\Z\bigr)
 =f(p-1)+s(p-1)=(f+s)(p-1).
\]
The preceding proof supplies the upper bound required here;
no assertion about its optimality is needed for the application.
Since a codeword weight is a sum of coordinate indicators, the
same estimate can be transferred through the coordinate functionals.
The next lemma carries this estimate to the divided weight residue.

\begin{lemma}\label{lem:degree}
Let $s$ be a nonnegative integer. If all weights of an
$\F_{p^f}$-linear code are divisible by $p^s$, then
\[
 F_s(u)=\frac{\wt(uG)}{p^s}\pmod p
\]
is a well-defined $\F_p$-valued function on its coefficient space.
It is scalar invariant, satisfies $F_s(0)=0$, and obeys
\[
 \deg_p F_s\le(f+s)(p-1).
\]
\end{lemma}

\begin{proof}
Let $V$ be the coefficient space, and let
$\lambda_i:V\to\F_q$ be its coordinate functionals.
With $R=\Z/p^{s+1}\Z$, define the nonzero indicator
\[
 \eta:\F_q\longrightarrow R,\qquad
 \eta(x)=\begin{cases}0,&x=0,\\1,&x\ne0.\end{cases}
\]
Then the weight function modulo $p^{s+1}$ is
\[
 W(u)=\sum_{i=1}^n\eta(\lambda_i(u))
       =\wt(uG)\pmod{p^{s+1}}.
\]
For $v\in V$, linearity of $\lambda_i$ gives the pullback
identity
\begin{align*}
 \delta_v(\eta\circ\lambda_i)(u)
 &=\eta(\lambda_i(u)+\lambda_i(v))
       -\eta(\lambda_i(u))\\
 &=\bigl(\delta_{\lambda_i(v)}\eta\bigr)(\lambda_i(u)).
\end{align*}
Iterating this identity and applying
Lemma~\ref{lem:additive-differences}, with
$L=(f+s)(p-1)+1$, yields
\begin{align*}
 \delta_{v_1}\cdots\delta_{v_L}W
 &=\sum_{i=1}^n
       \bigl(\delta_{\lambda_i(v_1)}\cdots
              \delta_{\lambda_i(v_L)}\eta\bigr)\circ\lambda_i\\
 &=0
 \qquad(v_1,\ldots,v_L\in V).
\end{align*}

The divisibility hypothesis allows this congruence to be
divided by $p^s$ in the integers. More explicitly, formula
\eqref{eq:iterated-difference} says that
\[
 \sum_{J\subseteq\{1,\ldots,L\}}(-1)^{L-|J|}
 \wt\left(\left(u+\sum_{j\in J}v_j\right)G\right)
 \equiv0\pmod{p^{s+1}}.
\]
Each summand is an integer multiple of $p^s$. Dividing all
summands by that integer gives
\[
 \sum_{J\subseteq\{1,\ldots,L\}}(-1)^{L-|J|}
 \frac{\wt\left(\left(u+\sum_{j\in J}v_j\right)G\right)}{p^s}
 \equiv0\pmod p.
\]
Equivalently,
\begin{equation}\label{eq:weight-residue-differences}
 \delta_{v_1}\cdots\delta_{v_L}F_s=0
 \qquad(v_1,\ldots,v_L\in V).
\end{equation}
This division does not invert $p$ in $R$. It uses the
additive isomorphism
\[
 p^sR\longrightarrow\F_p,\qquad
 p^sz\longmapsto z\pmod p,
\]
which is well defined because
$p^sz=p^sz'$ in $R$ is equivalent to $z\equiv z'\pmod p$.

To convert \eqref{eq:weight-residue-differences} into an
algebraic-degree bound, identify $V$ with $\F_p^m$ by a
prime-field basis. Write its reduced polynomial as
\[
 F_s(z)=\sum_{\boldsymbol{a}\in\{0,\ldots,p-1\}^m}
           c_{\boldsymbol{a}}z_1^{a_1}\cdots z_m^{a_m}.
\]
If $F_s=0$, the claimed bound is immediate. Otherwise let
$D$ be its degree, and choose a nonzero coefficient
$c_{\boldsymbol{a}}$ with $\sum_i a_i=D$. The elementary identity
\[
 (Z+1)^a-Z^a=\sum_{j=0}^{a-1}\binom aj Z^j
 \qquad(1\le a<p)
\]
shows that a difference in one coordinate lowers its exponent,
and $a$ such differences send $Z^a$ to $a!$.
For the coordinate
unit vectors $e_1,\ldots,e_m$, the difference operator
\[
 \delta_{e_1}^{a_1}\cdots\delta_{e_m}^{a_m}
\]
kills all monomials of degree less than $D$. It also kills
every other monomial of degree $D$, since such a monomial
has an exponent smaller than $a_i$ in some coordinate $i$.
For the selected monomial it gives
\[
 \delta_{e_1}^{a_1}\cdots\delta_{e_m}^{a_m}
     \bigl(c_{\boldsymbol{a}}z_1^{a_1}\cdots z_m^{a_m}\bigr)
 =c_{\boldsymbol{a}}\prod_{i=1}^m a_i!\ne0.
\]
Here $a_i<p$, so none of the factorials vanishes in $\F_p$.
If $D\ge L$, the same operator must kill $F_s$ by
\eqref{eq:weight-residue-differences}: its first $L$
differences already give zero. This contradiction proves
$D\le L-1=(f+s)(p-1)$.

Finally,
\[
 \wt((\gamma u)G)=\wt(\gamma(uG))=\wt(uG)
 \quad(\gamma\in\F_q^*),\qquad \wt(0G)=0.
\]
Therefore $F_s$ is scalar invariant and $F_s(0)=0$.
\end{proof}

\section{All layers in the basic valuation window}\label{sec:window}

The endpoint and degree lemmas supply the two ingredients needed
to apply the geometric bound to weight residues. The proof has two
distinct induction parameters: induction on $s$ supplies the divisibility
needed to define $F_s$, and minimality of the dimension supplies
its whole-zero hyperplanes. Making these roles explicit avoids
assuming the layer that is being proved. For $s=0$, the
conclusion is Ward's known $p$-divisibility result
\cite[Proposition 13]{Ward2001}; we retain this layer in the
uniform argument.

\begin{theorem}\label{thm:window}
For every integer $s$ with $0\le s<f$, a $q$-ary Griesmer code
with $p^{f+s}\mid d$ is $p^{s+1}$-divisible.
\end{theorem}

\begin{proof}
For a fixed field $\F_q$, let $\mathcal P_s$ be the assertion
\[
 p^{f+s}\mid d(C)
 \quad\Longrightarrow\quad
 p^{s+1}\mid\wt(c)\quad(c\in C)
\]
for every Griesmer code $C$, in every positive dimension.
We prove $\mathcal P_s$ successively for $s=0,\ldots,f-1$.
At $s=0$, all integer weights are divisible by $p^0=1$.
At $s>0$, assume $\mathcal P_{s-1}$. Since
\[
 p^{f+s}\mid d\quad\Longrightarrow\quad p^{f+s-1}\mid d,
\]
that preceding assertion makes every code under consideration
$p^s$-divisible. Thus the prerequisite for
Lemma~\ref{lem:degree} is available at every layer.

Suppose, for the current $s$, that $\mathcal P_s$ is false.
Choose a counterexample $C$ of smallest positive dimension
among all Griesmer codes with $p^{f+s}\mid d(C)$.
For its coefficient space $V=\F_q^k$, the function
\[
 F_s(u)=\frac{\wt(u)}{p^s}\pmod p
\]
is well defined. The failure of $p^{s+1}$-divisibility means
that it is nonzero. Lemma~\ref{lem:degree} also gives
\[
 F_s(0)=0,\qquad F_s(\gamma u)=F_s(u)\ (\gamma\in\F_q^*),
 \qquad\deg_p F_s\le(f+s)(p-1).
\]

The counterexample has $k\ge3$. Indeed, in dimension one
all nonzero words are scalar multiples of one another and
have weight $d$. In dimension two, the condition
$p^{f+s}\mid d$ implies $q\mid d$; the final assertion of
Lemma~\ref{lem:endpoints} again makes all nonzero weights
equal to $d$. In both dimensions,
$s+1\le f+s$ gives $p^{s+1}\mid d$, contradicting the
choice of $C$.

Now let $E\subseteq\PG(V^*)$ be the endpoint set of $C$.
For $P=\langle\lambda\rangle\in E$, the effective code
on $\ker\lambda$ has dimension $k-1$, minimum distance
$d$, and length $g_q(k-1,d)$, by
Lemma~\ref{lem:endpoints}. It still satisfies
$p^{f+s}\mid d$. Minimality of $k$ therefore gives
\[
 p^{s+1}\mid\wt(u)
 \quad(u\in\ker\lambda),\qquad
 F_s|_{\ker\lambda}\equiv0.
\]
Different projective points give different kernels: for
nonzero functionals $\lambda,\gamma$, equality
$\ker\lambda=\ker\gamma$ implies
$\Span(\lambda)=\Span(\gamma)$ by taking annihilators.
Thus $F_s$ has at least $|E|$ distinct whole-zero
hyperplanes. Theorem~\ref{thm:hyperplane} yields
\begin{equation}\label{eq:endpoint-upper}
 |E|\le p^s(q-1).
\end{equation}

The endpoint lower bound contradicts this upper bound.
In the notation of Lemma~\ref{lem:endpoints},
\[
 Q=q^{k-1}=p^{f(k-1)},\qquad
 f+s\le2f-1<2f\le f(k-1).
\]
Since $p^{f+s}$ divides both $d$ and $Q$, and $t>0$, we have
\[
 t=d-(\mu-1)Q\in p^{f+s}\Z_{>0}
 \quad\Longrightarrow\quad t\ge p^{f+s}.
\]
Consequently,
\[
 p^s(q-1)\ge |E|\ge g_q(k,t)\ge t
                  \ge p^{f+s}=p^sq>p^s(q-1),
\]
an impossibility. This proves $\mathcal P_s$ and completes
the induction on the layer.
\end{proof}

To record precisely what has now been established, put
$e=\nu_p(d)$. If $e<f$, the divisor in
Theorem~\ref{thm:ward} is $p^0=1$. If $f\le e<2f$, then
\[
 s=e-f\in\{0,\ldots,f-1\},\qquad
 p^{f+s}\mid d,
\]
and Theorem~\ref{thm:window} gives the exponent
$s+1=e-f+1$. We have therefore proved
Theorem~\ref{thm:ward} in every dimension throughout the
range $\nu_p(d)<2f$. The remaining argument transfers this
conclusion to all larger valuations.

\section{Transfer beyond the valuation window}\label{sec:transfer}

For $\nu_p(d)\ge2f$, a three-dimensional constant-weight
subcode provides representatives of every projective direction
in its dual. Ward's endpoint lifting lemma transfers those
directions to endpoints of the original code
\cite[Lemma 12]{Ward2001}. An elementary averaging argument
then produces a minimum-weight word with prescribed endpoint
ratios. The final projective-line identity and residual
induction follow the strategy of Ward's quaternary proof
\cite[Theorem 14]{Ward2001}; all steps are given below for
an arbitrary prime power $q$.

\subsection{A functional taking every field value}

The averaging step rests on the following criterion for a linear
functional to take every field value on a finite set. The criterion
concerns ordinary vectors, not projective points, and will be applied
after the endpoint representatives have been normalized.

\begin{lemma}\label{lem:onto}
Let $A$ be a set of $N>(q-1)^2$ distinct vectors in a
finite-dimensional $\F_q$-vector space $U$. There exists
$\varphi\in U^*$ such that $\varphi(A)=\F_q$.
\end{lemma}

\begin{proof}
Choose $\varphi$ uniformly from the finite vector space
$U^*$. For each $\alpha\in\F_q$, define the fiber size
\[
 n_\alpha(\varphi)=|\{v\in A:\varphi(v)=\alpha\}|,
 \qquad \sum_{\alpha\in\F_q}n_\alpha(\varphi)=N.
\]
For distinct $v,w\in A$, the vector $v-w$ is nonzero.
The evaluation map
\[
 U^*\longrightarrow\F_q,\qquad
 \varphi\longmapsto\varphi(v-w)
\]
is a surjective linear map and has a codimension-one kernel.
Thus
\[
 \Prob\bigl(\varphi(v)=\varphi(w)\bigr)=\frac1q
 \quad(v\ne w).
\]
For $v=w$, the probability is $1$. Counting ordered pairs
of elements of $A$ with equal images gives
\begin{align*}
 \E\sum_{\alpha\in\F_q}n_\alpha(\varphi)^2
 &=\sum_{(v,w)\in A^2}
       \Prob\bigl(\varphi(v)=\varphi(w)\bigr)\\
 &=N+\frac{N(N-1)}q.
\end{align*}

Suppose that every $\varphi$ omitted at least one value.
Then at most $q-1$ of its fibers would be nonempty.
Cauchy--Schwarz, applied to these nonempty fibers, would give
\[
 N^2
 =\left(\sum_{\alpha\in\F_q}n_\alpha(\varphi)\right)^2
 \le(q-1)\sum_{\alpha\in\F_q}n_\alpha(\varphi)^2.
\]
Taking expectations, multiplying by $q(q-1)$, and dividing
by $N>0$ yields
\begin{align*}
 \frac{N^2}{q-1}
 &\le N+\frac{N(N-1)}q,\\
 qN&\le(q-1)(N+q-1),\\
 N&\le(q-1)^2.
\end{align*}
This contradicts the hypothesis. Some functional therefore
takes all $q$ values on $A$.
\end{proof}

\subsection{Constant-weight sections and their endpoint fibers}

To use the preceding criterion, we obtain a set of vectors from
endpoint lifts of a constant-weight section. Repeated endpoint
shortening produces Griesmer sections of every smaller positive
dimension. When $q^2\mid d$, the three-dimensional sections are
constant-weight. This is also a consequence of
\cite[arXiv version, Theorem 4.6]{DHX}; their column multiplicities
can be computed directly by incidence counting.

\begin{lemma}\label{lem:lifts}
Let $C$ be a Griesmer code of dimension $k\ge3$, with
$q^2\mid d$. There exists a three-dimensional subspace
$D\le V$ whose effective code is Griesmer of distance $d$.
For every such $D$, all its nonzero words have weight $d$
and its projective column multiplicity is $d/q^2$ at every
point of $\PG(D^*)$. Moreover, every point of $\PG(D^*)$
has an endpoint lift in $E\subseteq\PG(V^*)$ under restriction
$V^*\to D^*$.
\end{lemma}

\begin{proof}
If $k=3$, take $D=V$. If $k>3$, apply
Lemma~\ref{lem:endpoints} repeatedly. At each stage an
endpoint exists, and its effective shortening lowers the
dimension by one while preserving the exact distance $d$
and the Griesmer equality. Thus there is a chain
\[
 V=D_k\supset D_{k-1}\supset\cdots\supset D_3=D,
 \qquad\dim D_j=j,
\]
whose effective codes have parameters
$[g_q(j,d),j,d]_q$. The restrictions remain subspaces of
the original coefficient space: restoring deleted zero
coordinates embeds each effective code into $C$.

Now consider any such three-dimensional $D$. Since
$q^{\dim D-1}=q^2\mid d$, the last assertion of
Lemma~\ref{lem:endpoints} gives
\[
 \wt(b)=d\quad(b\in D\setminus\{0\}),\qquad
 n_D=g_q(3,d)=d+\frac dq+\frac d{q^2}.
\]
Let $M_D$ be its column multiset in the projective plane
$\PG(D^*)$. Every line is the annihilator of a nonzero
$b\in D$, so
\[
 M_D(L)=n_D-d\quad\text{for every line }L.
\]
Fix a point $Q\in\PG(D^*)$. There are $q+1$ lines through
$Q$; each other point is on exactly one of them. Hence
\begin{align*}
 \sum_{L\ni Q}M_D(L)
 &=(q+1)M_D(Q)+\sum_{Q'\ne Q}M_D(Q')\\
 &=qM_D(Q)+n_D.
\end{align*}
Substituting the common line multiplicity gives
\[
 qM_D(Q)+n_D=(q+1)(n_D-d),
\]
and therefore
\begin{align*}
 M_D(Q)
 &=n_D-\frac{q+1}{q}d\\
 &=\left(d+\frac dq+\frac d{q^2}\right)
       -\left(d+\frac dq\right)
 =\frac d{q^2}.
\end{align*}
Thus every projective point of $D^*$ is an endpoint of
the effective code on $D$.

We next lift these points to endpoints of the original
code. The restriction map
\[
 \rho:V^*\longrightarrow D^*,\qquad
 \lambda\longmapsto\lambda|_D
\]
is surjective, since a linear functional on $D$ extends
to $V$ after extending a basis of $D$ to a basis of $V$.
Its kernel has dimension $k-3$. The associated projective
map is
\[
 \overline\rho:
 \PG(V^*)\setminus\PG(\ker\rho)\longrightarrow\PG(D^*),
 \qquad
 \langle\lambda\rangle\longmapsto\langle\rho(\lambda)\rangle.
\]
The excluded points represent exactly those functionals
that vanish on $D$, namely the zero coordinates deleted
from its effective code.

Fix $Q=\langle\beta\rangle\in\PG(D^*)$, with
$\beta\ne0$. Every point in the projective fiber
$\overline\rho^{-1}(Q)$ has a unique representative
$\lambda$ satisfying $\rho(\lambda)=\beta$: divide an
arbitrary representative by the unique nonzero scalar
relating its restriction to $\beta$. Conversely, distinct
elements of the affine fiber $\rho^{-1}(\beta)$ cannot
be proportional, because both restrict to the same
nonzero $\beta$. Therefore
\[
 \overline\rho^{-1}(Q)\longleftrightarrow\rho^{-1}(\beta)
 \quad\text{is a bijection},\qquad
 |\overline\rho^{-1}(Q)|=|\ker\rho|=q^{k-3}.
\]
Every original column in that projective fiber contributes
once to the multiplicity of $Q$ after restriction. Hence
\[
 M_D(Q)=\sum_{P\in\overline\rho^{-1}(Q)}M(P).
\]

Put $\mu=\lceil d/q^{k-1}\rceil$, and let
\[
 r_\beta=|E\cap\overline\rho^{-1}(Q)|.
\]
Each of these $r_\beta$ points has multiplicity $\mu$,
and every remaining point of the fiber has multiplicity
at most $\mu-1$. Consequently,
\begin{align*}
 \frac d{q^2}
 =M_D(Q)
 &\le\mu r_\beta
       +(\mu-1)\bigl(q^{k-3}-r_\beta\bigr)\\
 &=(\mu-1)q^{k-3}+r_\beta.
\end{align*}
With the endpoint remainder
$t=d-(\mu-1)q^{k-1}$, this becomes
\begin{equation}\label{eq:endpoint-fiber-lower}
 r_\beta\ge\frac d{q^2}-(\mu-1)q^{k-3}
            =\frac{t}{q^2}\ge1.
\end{equation}
The final inequality follows because $t>0$ and $q^2\mid t$:
both $d$ and $q^{k-1}$ are divisible by $q^2$.
This is the present three-dimensional instance of
\cite[Lemma 12]{Ward2001}. It proves that every $Q$ has
an original endpoint lift.
\end{proof}

\subsection{Completion of the divisibility proof}

The two preceding lemmas supply the remaining ingredient:
for a word outside every endpoint kernel, a minimum-weight
word can be chosen so that each of its affine combinations
with the given word enters an endpoint kernel. Those
combinations have known divisibility by dimension induction;
the residual code controls the remaining term in the
projective-line weight identity.

\begin{proof}[Proof of Theorem~\ref{thm:ward}]
Fix $q=p^f$. We use induction on the dimension, with the
claim in each dimension covering \emph{all} positive minimum
distances. Explicitly, the induction hypothesis at dimension
$k$ is that every Griesmer code $C'$ with $\dim C'<k$ is
\[
 p^{\max\{0,\nu_p(d(C'))-f+1\}}\text{-divisible}.
\]
The case $k=1$ is immediate, since every nonzero word has
weight $d$. Theorem~\ref{thm:window} and the observation
following its proof already cover every dimension when
$e=\nu_p(d)<2f$. It remains to consider
\[
 e\ge2f,\qquad q^2\mid d,\qquad
 \Delta=p^{e-f+1}.
\]
In dimensions two and three, $q^{k-1}\mid d$, so
Lemma~\ref{lem:endpoints} makes the code constant-weight.
Moreover,
\[
 e-f+1\le e,\qquad \Delta\mid p^e\mid d.
\]
These dimensions satisfy the desired conclusion. Assume
henceforth that $k\ge4$ and that the induction hypothesis
holds in every smaller dimension.

The induction hypothesis first settles the words in endpoint kernels.
Let $E$ be the endpoint set in $\PG(V^*)$. For
$P=\langle\lambda\rangle\in E$, the effective code on
$\ker\lambda$ is a dimension-$(k-1)$ Griesmer code with
the same minimum distance $d$. The induction hypothesis
therefore gives
\begin{equation}\label{eq:endpoint-word-divisibility}
 u\in\ker\lambda,\quad P\in E
 \quad\Longrightarrow\quad \Delta\mid\wt(u).
\end{equation}
This includes the zero word. Thus we need only treat a
word $u$ satisfying
\begin{equation}\label{eq:nonzero-on-endpoints}
 \lambda(u)\ne0
 \quad\text{for every }P=\langle\lambda\rangle\in E.
\end{equation}
Whether this condition holds is independent of the chosen
nonzero representative of $P$.

Choose a three-dimensional subspace $D\le V$ as in
Lemma~\ref{lem:lifts}. For each $Q\in\PG(D^*)$, choose
one original endpoint lift $P_Q\in E$ and a representative
$\lambda_Q\in V^*\setminus\{0\}$. Normalize its
restriction by its value at $u$:
\begin{equation}\label{eq:normalized-lifts}
 v_Q=\frac{\lambda_Q|_D}{\lambda_Q(u)}\in D^*.
\end{equation}
The denominator is nonzero by
\eqref{eq:nonzero-on-endpoints}. The numerator is nonzero
because $P_Q$ lies in a projective fiber above $Q$.
For $c\in\F_q^*$,
\[
 \frac{(c\lambda_Q)|_D}{(c\lambda_Q)(u)}
 =\frac{c(\lambda_Q|_D)}{c\lambda_Q(u)}=v_Q,
\]
so this normalization is independent of the representative.
Also
\[
 \langle v_Q\rangle=Q.
\]
Hence the vectors $v_Q$ are distinct for distinct $Q$.
Their set satisfies
\[
 \begin{aligned}
  A&=\{v_Q:Q\in\PG(D^*)\}\subseteq D^*,\\
  |A|&=\frac{q^3-1}{q-1}=q^2+q+1>(q-1)^2.
 \end{aligned}
\]

Apply Lemma~\ref{lem:onto} to $A$ in the vector space
$D^*$. It gives a functional $\varphi\in(D^*)^*$ with
$\varphi(A)=\F_q$. The natural evaluation isomorphism
\[
 D\longrightarrow(D^*)^*,\qquad
 b\longmapsto\bigl(\beta\mapsto\beta(b)\bigr)
\]
identifies this functional with some $b\in D$. Thus
\begin{equation}\label{eq:onto-ratios}
 \left\{\frac{\lambda_Q(b)}{\lambda_Q(u)}:
                Q\in\PG(D^*)\right\}=\F_q.
\end{equation}
Since $q\ge2$, a functional taking all $q$ values is not
zero, and therefore $b\ne0$. The constant-weight property
of $D$ yields
\[
 \wt(b)=d.
\]

For every $\alpha\in\F_q$, equation~\eqref{eq:onto-ratios}
provides a point $Q$ with
\[
 \frac{\lambda_Q(b)}{\lambda_Q(u)}=\alpha,
 \qquad \lambda_Q(b-\alpha u)=0.
\]
Because $P_Q\in E$, equation~\eqref{eq:endpoint-word-divisibility}
gives
\begin{equation}\label{eq:pencil-divisibility}
 \Delta\mid\wt(b-\alpha u)
 \qquad(\alpha\in\F_q).
\end{equation}
The value $0$ in \eqref{eq:onto-ratios} also shows that
$b$ vanishes at some endpoint, whereas $u$ vanishes at
none. Thus $b$ and $u$ are linearly independent: if
$b=cu$ with $c\ne0$, every endpoint value of $b$ would
be nonzero, and $c=0$ would contradict $b\ne0$.

The affine combinations have the required divisibility. The remaining
information comes from the residual code with respect to the
minimum-weight codeword $bG$. Lemma~\ref{lem:residual} gives
\[
 I=\supp(bG),\qquad C_b=\pi_I(C),\qquad
 \dim C_b=k-1,\qquad d(C_b)=d/q,
\]
and $C_b$ is Griesmer. The dimension induction applies
even though its distance has changed: the induction
claim covers all distances. Since
\[
 \nu_p(d/q)=e-f\ge f,
\]
every weight of $C_b$ is divisible by
\[
 p^{\nu_p(d/q)-f+1}=p^{e-2f+1}.
\]
For the residual weight
\[
 r=\wt(\pi_I(uG))
   =|\supp(uG)\setminus\supp(bG)|,
\]
we obtain
\begin{equation}\label{eq:residual-delta}
 p^{e-2f+1}\mid r,
 \qquad \Delta=p^f p^{e-2f+1}\mid qr.
\end{equation}
It remains to relate these residual weights to the
original word. The average weight equation on the
two-dimensional subcode $\iota_G(\Span(b,u))$ is
\cite[Section 1 and the proof of Theorem 14]{Ward2001}
\begin{equation}\label{eq:projective-line-weight}
 \wt(u)+\sum_{\alpha\in\F_q}\wt(b-\alpha u)
 =q(d+r).
\end{equation}
For completeness, let $b_i=\lambda_i(b)$ and
$u_i=\lambda_i(u)$. At each coordinate, the contributions
to its left side are given by
\[
\begin{array}{c|c|c|c}
 (b_i,u_i)&{\bf1}_{\{u_i\ne0\}}
 &\displaystyle\sum_{\alpha\in\F_q}
                   {\bf1}_{\{b_i-\alpha u_i\ne0\}}
 &\text{total}\\ \hline
 (0,0)&0&0&0\\
 b_i\ne0,\ u_i=0&0&q&q\\
 b_i=0,\ u_i\ne0&1&q-1&q\\
 b_i\ne0,\ u_i\ne0&1&q-1&q
\end{array}
\]
Indeed, if $u_i\ne0$, exactly one scalar
$\alpha=b_i/u_i$ gives a zero term; if $u_i=0$, every
term equals $b_i$. Equivalently, the coordinate identity is
\[
 {\bf1}_{\{u_i\ne0\}}
 +\sum_{\alpha\in\F_q}{\bf1}_{\{b_i-\alpha u_i\ne0\}}
 =q\,{\bf1}_{\{(b_i,u_i)\ne(0,0)\}}.
\]
Summing it over $i$ proves \eqref{eq:projective-line-weight},
because
\[
 |\supp(bG)\cup\supp(uG)|
 =|\supp(bG)|+|\supp(uG)\setminus\supp(bG)|=d+r.
\]

Finally, \eqref{eq:pencil-divisibility} makes every term
in the sum in \eqref{eq:projective-line-weight} divisible
by $\Delta$. Also $\Delta\mid d$ and
\eqref{eq:residual-delta} imply $\Delta\mid q(d+r)$.
Reducing that exact identity modulo $\Delta$ therefore gives
\[
 \wt(u)
 =q(d+r)-\sum_{\alpha\in\F_q}\wt(b-\alpha u)
 \equiv0\pmod\Delta.
\]
This treats every word satisfying
\eqref{eq:nonzero-on-endpoints}; all remaining words were
covered by \eqref{eq:endpoint-word-divisibility}.
The dimension induction is complete, and the theorem
holds for every positive dimension and every valuation
$\nu_p(d)$.
\end{proof}

\end{document}